\documentclass[%
pre,
nofootinbib,
onecolumn,
notitlepage,
11pt,
reprint,%
]{revtex4-1}

\usepackage{amsmath,amsfonts,amsthm,amssymb}
\usepackage{graphicx}                          
\usepackage{setspace}  
\usepackage{caption}
\usepackage{listings}
\usepackage{tabularx}
\usepackage{xcolor}
\usepackage{indentfirst}
\usepackage{subcaption}
\usepackage{overpic}

\usepackage{verbatim}

\usepackage{enumerate}

\usepackage{bm}

\usepackage[font=footnotesize,labelfont=bf]{caption}

\makeatletter
\@addtoreset{equation}{section}
\makeatother

\newtheorem{remark}{Remark}[section]

\newcommand\dd{\mathrm{d}}
\newcommand\pp{\partial}

\newcommand\x{\bm{x}}
\newcommand\uvec{\mathbf{u}}
\newcommand\X{\mathbf{X}}
\newcommand\y{\bm{y}}

\newcommand\n{{\bf n}}

\usepackage{mhchem}

\begin{document}

\title{On variational principles for polarization responses in electromechanical systems}

\author{Yiwei Wang}
\email{ywang487@iit.edu}
\author{Chun Liu}%
 \email{cliu124@iit.edu}
\affiliation{Department of Applied Mathematics, Illinois Institute of Technology, Chicago, IL 60616, USA.
}%


\author{Bob Eisenberg}
\email{beisenbe@rush.edu}
\affiliation{Department of Applied Mathematics, Illinois Institute of Technology, Chicago, IL 60616, USA.
}%
\affiliation{Department of Physiology and Biophysics, Rush University, 1750 W. Harrison, Chicago IL 60612.
}

\begin{abstract}
  Classical electrodynamics uses a dielectric constant to describe the polarization response of electromechanical systems to changes in an electric field. We generalize that description to include a wide variety of responses to changes in the electric field, as found in most systems and applications. Electromechanical systems can be found in many physical and biological applications, such as ion transport in membranes, batteries, and dielectric elastomers. 
  We present a unified, thermodynamically consistent, variational framework for modeling electromechanical systems as they respond to changes in the electric field; that is to say, as they polarize. This framework is motivated and developed using the classical energetic variational approach (EnVarA). The coupling between the electric part and the chemo-mechanical parts of the system 
  is described either by Lagrange multipliers or various energy relaxations. 
  The classical polarization and its dielectrics and dielectric constants appear as outputs of this analysis. The Maxwell equations  then become universal conservation laws of charge and current, conjoined to an electromechanical description of  polarization.
  Polarization describes the entire electromechanical response to changes in the electric field and can sometimes be approximated as a dielectric constant or dielectric dispersion. 
  \end{abstract}

\maketitle

\section{Introduction}


Electromagnetism is often described by Maxwell field equations that form a general and precise description of electrodynamics in the absence of matter, with only two parameters, both of which are true constants that can be measured directly by experiments and are found to be remarkably constant in a wide range of conditions. In the presence of matter, like dielectrics, things are more complex, because the field changes things that are charged and the charge changes the field \cite{oppenheimer1930note}. These interactions depend on the mechanical properties of the system, the distribution of charge and mass, and the Maxwell equations themselves.

Classical electrodynamics was based on a particularly simple idealized  model of electromechanical charge in insulating dielectrics. In the ideal linear dielectrics of the classical Maxwell Equations, interactions are particularly simple and described by a dielectric constant $\varepsilon{_r},$  a single real number. That classical model is, however, unable to adequately describe the complicated interaction between charge and field in most materials as measured recently \cite{barsoukov2005impedance, banwell1972fundamentals, crenshaw2013theory, eisenberg2019dielectric, eisenberg2015electrical, fiedziuszko2002dielectric, gudarzi2021self, kremer2002broadband, landau2013electrodynamics,rao2012molecular, raicu2015dielectric, sindhu2006fundamentals, steinfeld2012molecules, stuart2021infrared}. It should not be a surprise that a model adequate to deal with measurements available in the 1850's (typically on a time scale of a tenth of a second)  would need revision in the 2020's when time scales of $10^{-9}$s are commonplace in experiments and applications.

Other electromechanical systems (beyond insulating dielectrics) are even more complex because other forces---like diffusion and convection---come into play. Both diffusion and convection move charges, and so change electric fields, that in turn act on the charges. These complex electro-mechanical systems involving diffusion and transport play pivotal roles in  physical and biological applications. Examples include ion transport in biological cells and across biological membranes, in batteries and other electromechanical technology. Indeed, similar interactions of holes, electrons and fields underlie the semiconductor devices of our technology.  In all these systems, particles do not move independently. Interactions are of great importance.

One of the most important electro-mechanical systems is the transport of charged particles in dilute solutions, which is often described by a Poisson-Nernst-Planck (PNP) equation \cite{eisenberg2010energy}. The movement of charged particles is a mechanical process, involving diffusion and convection, but the motion of the charges changes their positions, forms an electrical current, and thus changes the electric field in the system, 
which in turn changes the motion of the charged particles themselves. 


In systems involving diffusion, the particles interact through the electric field and use concentration gradients to create a PNP system \cite{eisenberg1996computing, griffith2013electrophysiology}.  The classical PNP equation can be written as \cite{griffith2013electrophysiology}
\begin{equation}\label{PNP}
  \begin{aligned}
    & \frac{\pp c_i}{\pp t} = \nabla \cdot \left( D_i (\nabla c_i + \frac{qz_i}{k_B T} c_i \nabla \varphi ) \right) \\
    & - \Delta \varphi = \frac{1}{\varepsilon} (\sum_{i = 1}^n q z_i c_i + \rho_0(\x)),
  \end{aligned}
\end{equation}
where $c_i(\x, t)$ is the number density of the $i$-th species of ions, $\varphi(\x, t)$ is the electrostatic potential, $\rho_0(\x)$ represents the density of any immobile background charge, $q$ is the elementary charge, $z_i$ is the electric charge of one molecule of the $i$-th species,
and $\varepsilon$ is the permittivity that measures electric polarizability of the solution.
Here the effects of magnetic fields are totally neglected, which is only true when there are no time-dependent magnetic fields so ${\bf E} = - \nabla \varphi$ is curl-free.
The interaction between ions \footnote{Holes and electrons in semiconductors share many of the properties of ions in solutions.} and field is imposed through the Poisson equation. 

The understanding of electromechanical coupling is rather limited. It is even unclear whether the PNP type equation, which is a mechanical description for transportation of charged point particles, is consistent with the Maxwell field equations in general. 
The dielectrics of classical Maxwell electrodynamics---without diffusion---can be viewed as simple electromechanical systems in which the electric field changes the location of charge by a particularly simple rule. There exits a large literature developing variational theories for electromechanical and magnetomechanical coupling in a range of systems of this type (excluding diffusion for the most part) \cite{bustamante2009nonlinear, dorfmann2005nonlinear, ericksen2002electromagnetic,ericksen2007theory, eringen1963foundations, liu2013energy,jelic2006dissipative, maggs2012minimizing, mcmeeking2007principle, mehnert2016nonlinear, ogden2011mechanics, suo2008nonlinear, sprik2021continuum, vagner2021multiscale}. Inspired by these, we build a thermodynamically consistent variational description of general electromechanical systems and extend it to include diffusion. 

The framework is motivated and developed using the classical energetic variational approach (EnVarA) that allows consistent incorporation of other fields, e.g., reactions \cite{wang2020field} and even temperature \cite{liu2020brinkman}. 
We 
write explicit models of  electromechanical systems that change the distribution of charge (and mass) as the electric field changes including elastic, electroelastic, and diffusion forces.
The key point is to isolate material properties in the Maxwell equations and use the classical theories of mechanics and diffusion to describe those material properties, using the EnVarA functional formulation. In this paper, dynamics and fluctuation are imposed in the mechanical part only. 
The electrical part of electromechanical coupling is imposed through models of the system of interest, or by a Lagrance multiplier in a way less dependent on a specific model.
Imposing dynamics on the electrical part---or on both the electrical and mechanical parts of the system---appears possible, but leads to complexities beyond the scope of this paper.

The constitutive properties are separated from the Maxwell equations in this approach, allowing the Maxwell equations to be universal and exact, and the constitutive equations to describe (electro)material properties. Constitutive and Maxwell equations are joined by the energy variational process either as functionals or partial differential equations, with boundary conditions appropriate for the model system and setup of interest.  
As an illustration, we re-derive the classical PNP system in the proposed framework. 


\section{Preliminary}

\subsection{Mechanics: energy variational approach}
Mechanical systems can often be described by their energy and the rate of energy dissipation as in the energetic variational approach 
\cite{giga2018variational}. 
One of the simplest mechanical systems is a spring-mass system 
\begin{equation}\label{spring_mass}
  \begin{cases}
    & \x_t = {\bm v} \\
    & m {\bm v}_t = - \gamma \x_t  - \nabla V(\x),   \\
  \end{cases}
\end{equation}
where ${\bm v}$ is the velocity and $V(\x)$ is the potential energy. For a linear spring, $V(\x) = {\textstyle \frac{1}{2}} k |\x|^2$.
It is straightforward to show that the spring-mass system (\ref{spring_mass}) satisfies an energy-dissipation identity
\begin{equation}\label{FP_0} 
 \frac{\dd}{\dd t} \left( \frac{m}{2} |\x_t|^2 + V(\x) \right) = - \gamma |\x_t|^2,
\end{equation}
where $\mathcal{K} = \frac{m}{2} |\x_t|^2$ is the kinetic energy, $\mathcal{U} = V(\x)$ is the internal energy and $\gamma |\x_t|^2$ is the rate of energy dissipation due to the friction.

If the system also involves a stochastic force, modeled by a Gaussian white noise, then the dynamics becomes
\begin{equation}\label{Langevin}
  \begin{cases}
    & \x_t = {\bm v} \\
    & m {\bm v}_t = - \gamma v - \nabla V(\x) + {\bm \xi}(t),    \\
  \end{cases}
\end{equation}
where ${\bm \xi}(t)$ is a stochastic force satisfying $\langle {\bm \xi}, {\bm \xi}' \rangle = 2 k_B T m^{-1} \gamma \delta(t - t')$ due to the  fluctuation-dissipation
theorem (FDT) \cite{ma2016derivation}. The FDT ensures the system admits an energy-dissipation law and reaches the correct equilibrium state \cite{ma2016derivation}.
Here we adopt a Langevin representation, understanding fully well that this description is a constitutive model that needs to be confirmed by experiment and comparison with the actual properties of trajectories in matter and in accurate simulations of atomic motion.
Let $f(\x, {\bm v}, t)$ be the probability of a particle in location $\x$ with velocity ${\bm v}$, the Fokker-Planck equation of $f(\x, {\bm v}, t)$ corresponding to the Langevin dynamics (\ref{Langevin}) is given by
\begin{equation} \label{FP_1}
  \begin{aligned}
    & \pp_t f + \nabla_{\x} \cdot ( {\bm v} f) + \nabla_{\bm v} \cdot \left(( - \frac{\gamma}{m} {\bm v} - \frac{1}{m} \nabla V ) f \right) \\
    & = \frac{k_B T \gamma}{m^2} \Delta_v f. \\
  \end{aligned}
\end{equation}
Note that this is the full Langevin equation including the acceleration term.
Direct calculation reveals that the Fokker-Planck equation (\ref{FP_1}) for the full Langevin equation satisfies an energy-dissipation identity
\begin{equation}\label{ED_FP_1}
  \begin{aligned}
& \frac{\dd}{\dd t} \int  \frac{m}{2} f |v|^2  + k_B T   f  \ln f + V f \dd v \dd \x  \\
& = \int - \gamma f |{\bm v} + k_B T m^{-1}\nabla_{v} \ln f|^2 \dd {\bm v} \dd \x \leq 0. \\
  \end{aligned}
\end{equation}
The $k_B T f \ln f$ term comes from the noise, which corresponds to the entropy $T \mathcal{S}$ in classical thermodynamics. 

Much of the literature \cite{schuss1980singular, schuss2001derivation, eisenberg1995diffusion, grasser2003review, wu2015diffusion} follows Smoluchowski and Einstein and is concerned with overdamped systems.  Many applications occur in highly overdamped systems, like ionic solutions or liquids. These are condensed phases, with almost zero empty space. In such systems, atoms cannot move without strong interactions (`collisions') which become frictional and dissipative after a very short time, of the order of $10^{-14}$s. This literature derives and discusses the over damped Langevin equation from the perspective of the theory of stochastic processes \cite{schuss1980singular, schuss2001derivation, eisenberg1995diffusion} or the Boltzmann transport integral \cite{eisenberg1995diffusion, grasser2003review, wu2015diffusion} while our perspective is energetic. Each treatment uses slightly (but significantly) different definitions of `overdamped', `flux', and 'mean velocity'. We are no exception. If definitions are assumed to be identical in these different approaches, confusion can result.

In an overdamped region ($\gamma \gg m$), the inertial term in (\ref{Langevin}) can be ignored and the dynamics can be reduced to an overdamped Langevin equation (after rescaling but keep the same notation)
\begin{equation} \label{Smoluchowski}
 \gamma \x_t = - \nabla V(\x) + {\bm \xi}(t),  
\end{equation}
where $\langle {\bm \xi}, {\bm \xi}' \rangle = 2 k_B T \gamma \delta(t - t')$.
The corresponding Fokker-Planck equation of $\rho(\x, t)$ then becomes
\begin{equation}\label{FP_2}
\rho_t = \nabla \cdot \left( \frac{1}{\gamma} (k_B T \nabla \rho + \rho \nabla V ) \right),
\end{equation}
where $\rho(\x, t)$ is the probability distribution of finding the particle at location $\x$,
If we define the average velocity as
\begin{equation} \label {average velocity}
\uvec = \frac{1}{\gamma} \nabla  \left( k_B T (\ln \rho + 1) +  V \right),
\end{equation}
then energy-dissipation law of the Fokker-Planck equation (\ref{FP_2}) can be formulated as
\begin{equation}
\frac{\dd}{\dd t} \int( (k_B T)( \rho \ln \rho) + V(\x))\, \dd \x = - \int \gamma \rho |\uvec|^2 \,\dd \x.
\end{equation}
Again, the $(k_BT)( \rho \ln \rho)$ term corresponds to $- T \mathcal{S}$ with $\mathcal{S} = - \rho \ln \rho$ being the entropy.

In general, as in previous examples, an isothermal mechanical system can be well defined through an energy-dissipation law
\begin{equation}\label{Eenergy_Law}
\frac{\dd}{\dd t} (\mathcal{K} + \mathcal{F}) = - \triangle,
\end{equation}
along with the kinematics of the employed variables.
Here $\mathcal{K}$ is the kinetic energy, $\mathcal{F} = \mathcal{U} - T \mathcal{S}$ is the Helmholtz free energy, and $\triangle$ is the rate of the energy dissipation, which is the entropy production in the system \cite{giga2018variational}.
From the energy-dissipation law (\ref{ED1}), the corresponding evolution equation can be derived by the energetic variational approach (EnVarA). 

In more detail: EnVarA consists of  two distinct variational processes: the Least Action Principle (LAP) and the Maximum Dissipation Principle (MDP) \cite{giga2018variational}. The LAP states that the dynamics of a Hamiltonian system are determined as a critical point of the action functional $\mathcal{A}(\x) = \int_0^T (\mathcal{K} - \mathcal{F})\, \dd t$ with respect to $\x(\X, t)$ (the trajectory for mechanical systems, where $\X$ are Lagrangian coordinates) \cite{giga2018variational}, i.e., 
\begin{equation}
  \delta \mathcal{A} =  \int_{0}^T \int_{\Omega(t)} (f_{\text{inertial}} - f_{\text{conv}})\cdot \delta \x~  \dd \x  \, \dd t.
\end{equation}
The dissipative force in such a system can be determined by minimizing the dissipation functional $\mathcal{D} = \frac{1}{2} \triangle$ with respect to the ``rate'' $\x_t$ in the linear response regime \cite{de2013non}, i.e.,
\begin{equation}
\delta \mathcal{D} = \int_{\Omega(t)} f_{\text{diss}} \cdot \delta \x_t~ \dd \x.
\end{equation}
This principle is known as Onsager's MDP \cite{onsager1931reciprocal, onsager1931reciprocal2}.
According to force balance, which is Newton's second law if we view inertial force as $ma$, we have, in Eulerian coordinates,
\begin{equation}\label{FB}
\frac{\delta A}{\delta \x} = \frac{\delta \mathcal{D}}{\delta \x_t}
\end{equation}
This describes the dynamics of the system. It is worth mentioning that in principle, mechanical systems are totally determined by the trajectory or the flow map $\x(\X, t)$, as indicated by the above variational procedure.
To further illustrate the framework of EnVarA, we consider a simple class of mechanical process, generalized diffusions, which are concerned with the evolution of a conserved quantity $c(\x, t)$ satisfying the kinematics (conservation of mass)   
\begin{equation}\label{kinematics}
\pp_t c + \nabla \cdot (c \uvec) = 0,
\end{equation}
where $\uvec$ is an average velocity. 

In the framework of EnVarA, a diffusion process 
can be described by the energy-dissipation law
\begin{equation}
\frac{\dd}{\dd t} \int \omega(c) \,\dd \x = - \int \eta(c) |\uvec|^2 \,\dd \x,
\end{equation}
where $\omega(c)$ is the free energy, $\eta(c)$ is the friction coefficient. 
Then a standard variational process leads to a force balance equation \cite{giga2018variational, liu2020lagrangian}
\begin{equation}\label{FB1}
  \eta(c) \uvec =  - \nabla \left( \frac{\pp \omega}{\pp c} c - \omega(c) \right) = - c \,\nabla \mu,
\end{equation}
where $$\mu = \frac{\pp \omega}{\pp c}~ \text{is the chemical potential}.$$ The complete derivation of (\ref{FB1}) can be found in Appendix A for self-consistency.

A typical example of $\omega(c)$ is
\begin{equation*}
\omega(c) = (k_B T)\, c \ln c +  {\textstyle \frac{1}{2} \int K(\x, \y) \,c(\y)c(\x)\, \dd \y},
\end{equation*}
where the first term is the entropy, and the second term is the internal energy that models the interaction between particles, with $K(\x, \y)$ being the interaction kernel. The interaction can include a steric potential as well as the Coulomb potential \cite{liu2020molecular}.
In that way, the interactions can include forces arising from the finite size of ions that limit the total concentration of ions to a finite number producing the saturation phenomena so characteristic of biology. 
Then the variational procedure leads to expressions like $c \uvec =  -  (k_B T \,\nabla c + c\, \nabla (K * c)).$ Combing the force balance equation (\ref{FB1}) with the kinematics (\ref{kinematics}) and taking $\eta(c) = c$ (for simplicity), we obtain a non-local diffusion equation
\begin{equation}
c_t = \nabla \cdot (k_B T \, \nabla c + c \,\nabla (K * c)).
\end{equation}

Similarly, the PNP equation (\ref{PNP}) can be derived from the energy-dissipation law \cite{eisenberg2010energy}
\begin{equation}\label{ED_PNP}
\frac{\dd}{\dd t} \int \sum_{i=1}^n k_B T\, c_i (\ln c_i - 1) + \frac{\varepsilon}{2} | \nabla \varphi|^2 = - \int \frac{k_B T}{D_i} c_i |{\uvec_i}|^2 \dd \x
\end{equation}
with the constraint
\begin{equation}\label{Poisson}
     - \nabla \cdot ( \varepsilon (\nabla \varphi) ) = (\sum_{i = 1}^n q z_i c_i + \rho_0(\x)),
\end{equation}
which is a differential form of the Gauss's law. We refer the interested readers to \cite{eisenberg2010energy} for detailed derivation.
However, this formulation assumes the existence of a dielectric constant $\varepsilon$, as well as the electric potential $\varphi$ in advance. Moreover, as proposed in \cite{de2013non,  dreyer2016new, muller1985}, the proper thermodynamic variable for a thermodynamically consistent description of electrodynamics is the electric field ${\bf E}$ (or ${\bf D}$), rather than the electrostatic potential $\varphi$.

\subsection{Electricity: Maxwell field equations in vacuum}
The fundamental equations in classical electromagnetism are Maxwell's field equations, which can be formulated as
\begin{equation}\label{Maxwell-vacuum}
  \begin{aligned}
    & \nabla \cdot (\varepsilon_0 {\bf E} ) =   0 \\
    & \nabla \cdot {\bf B} = 0 \\
    & \frac{\pp {\bf B}}{\pp t} = - (\nabla \times {\bf E}) \\
  &  \nabla \times {\bf B} = \mu_0 \left( \varepsilon_0 \dfrac{{\pp \bf E}}{\pp t} \right),
  \end{aligned}
\end{equation}
in vacuum, where ${\bf E}$ and ${\bf B}$ are electric and magnetic field, $\varepsilon_0$ is the electrical constant also called the permittivity of free space and $\mu_0$ is the magnetic constant, also called  the permeability of free space. 

Direct calculations show the Maxwell equations (\ref{Maxwell-vacuum}) satisfy the  energy-dissipation law 
\begin{equation}\label{ED1}
  \begin{aligned}
 & \frac{\dd}{\dd t} \int_{\Omega}(  \frac{\varepsilon_0}{2}   |{\bf E}|^2 +  \frac{1}{2 \mu_0} |{\bf B}|^2) \,\dd x 
=   -  \int   \frac{1}{\mu_0} \nabla \cdot ({\bf E} \times {\bf B} ) \,\dd \x, \\
&  \quad  =  - \int_{\pp \Omega}    \frac{1}{\mu_0}  ({\bf E} \times {\bf B}) \cdot {\bm \nu} \;\dd S. \\
  \end{aligned}
\end{equation}
Motivated by the above calculation, we can define the electric field energy density $e_F({\bf E}, {\bf B})$ as
\begin{equation}\label{ef_EB}
  e_F({\bf E}, {\bf B}) = \frac{\varepsilon_0}{2} |{\bf E}|^2 + \frac{1}{2 \mu_0} |{\bf B}|^2.
\end{equation}
The vector $\frac{1}{\mu_0} {\bf E} \times {\bf B}$ is the Poynting vector that represents the directional energy flux (the energy transfer per unit area per unit time) of an electromagnetic field.

Conventionally, one introduces the electric and magnetic displacement vectors ${\bf D}$ and ${\bf H}$, defined by
\begin{equation}\label{aether}
{\bf D} = \varepsilon_0 {\bf E}, \quad {\bf H} = \frac{1}{\mu_0} {\bf B}.
\end{equation}
Relations (\ref{aether}) are known as Lorentz-Maxwell {\ae}ther relations.
It can be noticed that
\begin{equation}\label{DH_vacuum}
{\bf D}= \frac{\pp e_F}{\pp {\bf E}}, \quad {\bf H} = \frac{\pp e_F}{\pp {\bf B}},
\end{equation}
which provides an energetic variational formulation for ${\bf D}$ and ${\bf E}$ connecting the energies and these classic fields.

\begin{remark}
The field energy $e_{F}$ per unit volume can be formulated by choosing different primitive variables. For instance, in \cite{ericksen2007theory}, the field energy $e_{F}$ is defined as
\begin{equation}\label{ef_BD}
 e_F({\bf B}, {\bf D}) = \frac{ |{\bf B}|^2}{2 \mu_0} + \frac{|{\bf D}|^2}{2 \varepsilon_0}
\end{equation}
by using ${\bf B}$ and ${\bf D}$ as independent variables. The electric and magnetic field ${\bf E}$ and ${\bf H}$ can be defined as
\begin{equation}
{\bf E} = \frac{\pp e_F}{\pp {\bf D}} = \frac{1}{\varepsilon_0} {\bf D}, \quad {\bf H} = \frac{\pp e_F}{\pp {\bf B}} = \frac{1}{\mu_0} {\bf H}.
\end{equation} We refer the interested readers to \cite{bustamante2009nonlinear} for detailed discussions on different variational formulations. 
\end{remark}


The simplest electromechanical system has a point charge in the electric field. In general, a charge in the electric field is not only subjected to a force exerted by the field, but also changes the field in turn.  In such systems the electric field must be computed from the charges, in models combining electrical and mechanical theories, as we do here \cite{eisenberg1996computing}.

The electric potential can be constant and independent of a charge at a particular location if it is `voltage clamped' by an experimental apparatus that supplies charge and energy as in the classical voltage clamp systems of membrane biophysics.
The electric field can be similarly constant only if many potentials, at many points are each separately voltage clamped by their own apparatus. We are unaware of experiments that do this, see \cite{han1993superconducting}.

For a particle of charge  $q_k$ and velocity ${\bm v}_k$, the \emph{Lorentz} force on the particle is given by
\begin{equation}
  f = q_k ({\bf E} + {\bm v}_k \times {\bf B}).
\end{equation}
Then the movement of the particle can be described by
\begin{equation}
 m \ddot{\x}_k =  
 q_k ({\bf E} + {\bm v}_k \times {\bf B}),
\end{equation}
where ${\bm v}_k = \dot{\bm x}_k$ is the velocity of the particle.
It is easy to show the following energy identity
\begin{equation}\label{Kinetic_E}
\tfrac{\dd}{\dd t} ( {\textstyle \frac{1}{2}} m {\bm v}_k^2) = q_k {\bf E} \cdot {\bm v_k}  
\end{equation}
since $({\bm v}_k \times {\bf B}) \cdot {\bm v}_k = 0$.

The Maxwell field equation in this case can be formulated as 
\begin{equation}\label{Maxwell_vacuum_1}
  \begin{aligned}
    & \nabla \cdot (\varepsilon_0 {\bf E} ) =   \rho \\
    & \nabla \cdot {\bf B} = 0 \\
    & \frac{\pp {\bf B}}{\pp t} = - (\nabla \times {\bf E}) \\
  &  \nabla \times {\bf B} = \mu_0 \left( \varepsilon_0 \dfrac{{\pp \bf E}}{\pp t} + {\bf j} \right),
  \end{aligned}
\end{equation}
where the charge density $\rho$ and the (particle) current density ${\bm j}$ \cite{landau2013electrodynamics, eisenberg2017dynamics}  is defined by 
\begin{equation}
\rho = q_k \delta(\x - \x_k), \quad {\bm j} = q_k v_k \delta(\x - \x_k) = \rho {\bm v}_k,
\end{equation}
with $\delta$ being the Dirac delta function. Here, we assume that placing a charged particle in a vacuum does not change the Maxwell equations themselves.

It is straightforward to show that the equation (\ref{Maxwell_vacuum_1}) satisfies the following energy identity
\begin{equation}\label{ED11}
  \begin{aligned}
 & \frac{\dd}{\dd t} \int_{\Omega}  \frac{\varepsilon_0}{2} |{\bf E}|^2 + \frac{1}{2 \mu_0} |{\bf B}|^2) \,\dd x \\
& \quad = \int_{\Omega} - {\bf E} \cdot {\bm j} \;\dd \x  - \int_{\pp \Omega} \frac{1}{\mu_0}  ({\bf E} \times {\bf B}) \cdot {\bm \nu}\, \dd S. \\
  \end{aligned}
\end{equation}
Combining (\ref{Kinetic_E}) with (\ref{ED11}), we obtain the energy-dissipation law of the total electromechanical system
\begin{equation}
\frac{\dd}{\dd t} \int \frac{m}{2}  |{\bm v}_k|^2 \delta (\x - \x_k) +  \frac{\varepsilon_0}{2}  |{\bf E}|^2 +   \frac{1}{2 \mu_0} |{\bf B}|^2 = 0. 
\end{equation}
We can define $ c = \delta(\x - \x_k)$ as the {\bf number density} of the charged particles, then formally, the energy-dissipation law can be written as
\begin{equation}
    \frac{\dd}{\dd t} \int (\frac{m}{2} c |\,{\bm u}|^2  + \frac{\varepsilon_0}{2}  |{\bf E}|^2 +  \frac{1}{ 2 \mu_0} |{\bf B}|^2 ) \;\dd \x = 0  
\end{equation}
where $u(\x_k) =  {\bm v}_k$.

\begin{remark}
Similarly, if a charge particle is placed in a medium and satisfies the ordinary differential equation (the force balance between the mechanical force and the Lorentz force on the particle)
\begin{equation}
m \ddot{\x}_k + \gamma \dot{\x}_k + \nabla V(\x_k) = q_k ({\bf E} + \dot{\x}_k \times {\bf B}),
\end{equation}
then the energy-dissipation law is formally given by
\begin{equation}
  \begin{aligned}
  & \frac{\dd}{\dd t} \int \left(\frac{m c(\x) }{2} \,|{\bm u}|^2 + c(\x) V(\x)  + W ({\bf E}, {\bf B})\, \right) \dd \x \\
  & \quad = - \int \gamma  c(\x) |\uvec|^2 \dd \x, \\
  \end{aligned}
\end{equation}
where $c(\x) = \delta(\x - \x_k)$, $u(\x_k) = {\bm v}_k$, $V(\x)$ is the potential energy and $W({\bf E}, {\bf B})$ is the electromagnetic field energy in such a medium. The formulation also works for the case with $N$-particles although one must be careful in evaluating interactions of the different particles.
\end{remark}

\section{Variational treatment of electro-mechanical systems}


Motivated by the calculations in the last section, we present a general framework for deriving a thermodynamically consistent model involving electromechanical coupling by using an energetic variational approach.
The EnVarA framework allows a general treatment of the response of an electromechanical system to a change of the electric field. It includes classical polarization, even with complex time dependence and the classical model of an ideal dielectric. It also includes electromechanical systems that involve diffusion and translation, and other energy sources not present in the classical Maxwell equations. This framework makes minimal assumptions about the electric displacement field and the properties of its polarization component.

The framework starts with a general electromechanical free energy 
\begin{equation}
\mathcal{F}({\bf E}, {\bm \zeta}) = \int W({\bf E}, {\bm \zeta}) \dd \x ,
\end{equation}
where $W({\bf E}, {\bm \zeta})$ is the electromechanical free energy per unit volume, ${\bf E}$ is the electric field and ${\bm \zeta}$ represents other mechanical variables, such as densities of ions, the deformation tensor,  order parameters in liquid crystals.

We can generalize the definition of electric displacement field ${\bf D}$ in vacuum (\ref{DH_vacuum}) and define ${\bf D}$ as
\cite{bustamante2009nonlinear, landau2013electrodynamics, liu2013energy, suo2008nonlinear}
\begin{equation}\label{Def_D_from_E}
{\bf D} = \frac{\pp W({\bf E}, {\bm \zeta})}{\pp {\bf E}}.
\end{equation}
Consequently,  the electric polarization field ${\bf P}$ is defined by \cite{eringen1963foundations}
\begin{equation}
{\bf P} = {\bf D} - \varepsilon_0{\bf E}.
\end{equation}
So both ${\bf D}$ and ${\bf P}$ are derived from the electromechanical free energy $W({\bf E}, {\bm \zeta})$. 

The different  $W({\bf E}, {\bm \zeta})$ correspond to different constitutive relations between ${\bf D}$ and ${\bf E}$. For a `linear' dielectric, we have
\begin{equation}
  \mathcal{F}({\bf E}, {\bm \zeta}) = \omega({\bm \zeta}) + \frac{\varepsilon({\bm \zeta})}{2} |{\bf E}|^2
\end{equation}
Then 
\begin{equation}
{\bf D} = \varepsilon ({\bm \zeta}) {\bf E}
\end{equation}
and $\varepsilon ({\bm \zeta}) = \varepsilon_0 \varepsilon_r (\bm \zeta)$ is the conventional permittivity and $\varepsilon_r (\bm \zeta)$ is the dielectric constant. The form of free energy $W({\bf E}, {\bm \zeta})$ can be obtained from experiments by solving some inverse problems or from more-detailed model \cite{martin2020effect, zhuang2021like}. 
In general, the relation between ${\bf D}$ and ${\bf E}$ can be fully nonlinear.

As an illustration, let us first consider dielectric fluids. We can take $\rho$ and ${\bf E}$ as the state variables, and assume the free energy is given by \cite{sprik2021continuum}
\begin{equation}\label{Energy_E_rho}
\mathcal{F}(\rho, {\bf E}) = \mathcal{F}_M (\rho) + \mathcal{F}_{\rm elec} (\rho, {\bf E}),
\end{equation}
where $\mathcal{F}_M (\rho) = \int \omega(\rho) \dd \x$ is a purely mechanical component of the free energy, i.e., the free energy of the system in absence of the electric field ${\bf E}$. $\mathcal{F}_{\rm elec} (\rho, {\bf E})$ is the electromechanical energy, which is assumed to be
\begin{equation}
  \mathcal{F}_{\rm elec} (\rho, {\bf E}) = \frac{\varepsilon(\rho)}{2} |\bf E|^2,
\end{equation}
for linear dielectrics. Then, the variational procedure (\ref{Def_D_from_E}) leads to ${\bf D} = \varepsilon(\rho) {\bf E}$ and ${\bf P} = {\bf D} - \varepsilon_0 {\bf E}$.
For the pure dielectric case without any free charges, we have
\begin{equation}\label{cons_DE}
\nabla \cdot {\bf D} = 0, \quad \nabla \times {\bf E} = 0,
\end{equation}
which indicates that there exists an electrostatic potential $\varphi$ such that
\begin{equation}
{\bf E} = - \nabla \varphi
\end{equation}
and $\varphi$ satisfies the Poisson equation
\begin{equation}
- \nabla \cdot (\varepsilon(\rho) \nabla \varphi ) = 0.
\end{equation}

  The electrostatic potential $\varphi$ can be viewed as a Lagrange multiplier for the constraint $\nabla \cdot {\bf D} = 0$, along with the quasi-equilibrium condition, i.e., ${\bf E}$ minimize the free energy (\ref{Energy_E_rho}) without delay \cite{landau2013electrodynamics}. Indeed, we can introduce a Lagrange multiplier $\varphi$ for the constraint $\nabla \cdot {\bf D} = 0$, which leads to
  \begin{equation}\label{Energy_E_rho_lambda}
    \mathcal{F}(\rho, {\bf E}; \varphi) = \mathcal{F}_M (\rho) + \frac{\varepsilon(\rho)}{2} |{\bf E}|^2- \varphi (\nabla \cdot (\varepsilon(\rho) {\bf E})).
    \end{equation}
 By assuming that the electric field reaches equilibrium without delay, we have
    \begin{equation}
0  = \frac{\delta \mathcal{F}(\rho, {\bf E}; \varphi)}{\delta {\bf E}} = \varepsilon(\rho) {\bf E} + \varepsilon(\rho) \nabla \varphi,
    \end{equation}
    which leads to
    \begin{equation}
{\bf E} = - \nabla \varphi,
    \end{equation}
    and $\nabla \times {\bf E} = 0$.
    Delays will introduce additional dispersions into the impedance response, which will be discussed in future work.

Next we discuss the dynamics of the system, which is described by a suitable dissipation functional on the mechanical part. A simple choice of the dissipation is 
\begin{equation}
  \triangle = 2 \mathcal{D} = \int \eta(\rho) |\uvec|^2 \dd \x,
\end{equation}
where $\eta(\rho)$ is the friction coefficients.
By a standard variational procedure (see Appendix A), we have
\begin{equation}
- \eta(\rho) \uvec = \rho \nabla (\omega' (\rho) + \frac{1}{2} \varepsilon'(\rho) |{\bf E}|^2 ),
\end{equation}
which is equivalent to the results in \cite{landau2013electrodynamics} [page 68, eq. (15.12)].


For systems involving free charges, the electric displacement field ${\bf D}$ satisfies a differential version of Gauss's law
\begin{equation}\label{constraint_1}
 \nabla \cdot {\bf D} = \rho_{f}(\x),
\end{equation}
where $\rho_f(\x)$ is the total (electric) free charge density at $\x$.
From the mechanical part of the system, one can calculate the (mechanical) charge density $\hat{\rho}_f({\bm \zeta})$. 
However, in general, $\widehat{\rho}_f({\bm \zeta})$ may not be exactly the same as $\rho_f$ in the Gauss's law. For instance, polarization arises from the separation of the centers of positive and negative charges, which produces a difference between the mechanical charge density $\hat{\rho}_f({\bm \zeta})$ and the electric charge density $\rho_f(\x)$. In the energetic variational formulation, $\rho_f(\x)$ and $\hat{\rho}_f({\bm \zeta})$ can be linked by either a Lagrange multiplier or various energy relaxations.  Energy relaxations are a general way of describing electromechanical coupling.

In the following, we illustrate both approaches by modeling the transportation of charged particles in dilute solutions. We assume the free energy is given by
\begin{equation}
\mathcal{F}(c, {\bf E}) = \int \sum_{i=1}^n K_B T c_i (\ln c_i - 1) +  W_{\rm elec}({\bm c}, {\bf E}) \dd \x,
\end{equation}
and ${\bf D} = \frac{\pp W_{\rm elec} }{\pp {\bf E}}$.
From the mechanical part of the system, one can calculate the (mechanical) charge density as
\begin{equation}
\widehat{\rho}_f({\bm c}) = \sum_{i=1}^n q z_i c_i + \rho_0(\x), 
\end{equation}
$\rho_0(\x)$ is the density of any immobile background charge, $z_i$ is the electric variance of $i$-th species, and $q$ is the elementary charge. Both approaches can lead to PNP type systems with suitable dissipations.

\subsection{PNP equation with a Lagrange multiplier} 


In the first approach, we can introduce a Lagrange multiplier to link  $\widehat{\rho}_f({\bm c})$ and $\rho_f(\x)$ \cite{brenier2021relaxed, landau2013electrodynamics, jadhao2013free, maggs2012minimizing}, i.e.,
\begin{equation}
  \begin{aligned}
    & \mathcal{F}({\bm \rho}, {\bm E}; \varphi)  = \int_{\Omega}  \sum_{i=1}^n K_B T c_i (\ln c_i - 1) +    W_{\rm elec}({\bm c}, {\bf E})   \\
    & \quad - \varphi  \left( \nabla \cdot \left( \frac{\pp W_{\rm elec}}{\pp E} \right) - \sum_{i=1}^n q z_i c_i - \rho_0(\x) \right) \dd \x, \\
  \end{aligned}
\end{equation}
where $\varphi$ is the Lagrange multiplier.
In the situation when electric part can reach equilibrium without delay, 
we will have $\frac{\delta \mathcal{F}}{\delta {\bf E}} = 0.$
We are aware that introducing a delay will add a dispersion to the impedance response we calculate. Adding multiple delays of various types is likely to create most of the dispersions seen in the impedance literature \cite{barsoukov2005impedance, banwell1972fundamentals,  gudarzi2021self, kremer2002broadband, rao2012molecular, raicu2015dielectric, sindhu2006fundamentals, steinfeld2012molecules, stuart2021infrared}.

To derive the classical PNP equation (\ref{PNP}), we take $W_{\rm elec} ({\bf E})$  as
\begin{equation}
W_{\rm elec} = \frac{\varepsilon}{2} |{\bf E}|^2,
\end{equation}
with $\varepsilon$ being a constant, and ${\bf D} = \varepsilon {\bf E}$.
In the case, $\frac{\delta \mathcal{F}}{\delta {\bf E}} = 0$ leads to
\begin{equation}
  {\bf E} = - \nabla \varphi
  \end{equation}
The Lagrange multiplier $\varphi$ is the usual electrostatic potential $\varphi$.  

Next we look at the dynamics of the system, which are on mechanical part only. As we did in eq.(\ref{ED_PNP}), we can impose the dissipation as
\begin{equation}\label{Diss_PNP}
\triangle =  2\mathcal{D} = \int \sum_{i=1}^n \frac{k_B T}{D_i} c_i |\uvec_i|^2 \dd \x,
\end{equation}
where $\uvec_i$ is the average velocity of $i$-th species.
By a standard energetic variational approach (see Appendix A), we obtain
\begin{equation}
\frac{k_B T}{D_i} c_i \uvec = - c_i \nabla \mu_i
\end{equation}
where
\begin{equation}
\mu_i = \frac{\delta \mathcal{F}}{\delta c_i} = k_B T \ln c_i + \varphi c_i q.  
\end{equation}
The final PNP equation can be written as
\begin{equation}
  \begin{aligned}
    & \pp_t c_i = \nabla \cdot \left( D_i (\nabla c_i + \frac{q z_i}{k_B T}  c_i  \nabla \varphi  ) \right) \\
    & - \nabla \cdot (\varepsilon \nabla \varphi  ) = \sum_{i=1}^n q z_i c_i + \rho_0 (\x). \\
  \end{aligned}
\end{equation}

Although many studies assume that there exists a dielectric constant in ionic solution, the dielectric constant usually depends strongly on time and is typically  heterogenous. It often depends on other mechanical variables, such as the concentration of ions \cite{gavish2016dependence}. 
Within the above framework, it is easy for us to derive a PNP equation with heterogenous dielectric properties. As an example, consider a free energy
\begin{equation}
\mathcal{F}({\bm c}, {\bf E}) = \int \sum_{i=1}^N c_i (\ln c_i - 1) + \frac{\varepsilon({\bm c})}{2}  |{\bf E}|^2 \,\dd \x,  
\end{equation}
where $ \varepsilon ({\bm c}) = \varepsilon_0 \varepsilon_r ({\bm c})$ is the concentration-dependent permittivity.
In this case, the electric displacement vector is given by
\begin{equation}
{\bf D} =   \varepsilon ({\bm c}) {\bf E}.
\end{equation}
The Lagrange multiplier approach and the quasi-equilibrium assumption for the electric field ${\bf E}$ leads to
\begin{equation}
\frac{\delta \mathcal{F}}{\delta {\bf E}} 
= \varepsilon ({\bm c}) {\bf E} + \varepsilon ({\bm c}) \nabla \varphi = 0.
\end{equation}
We still have ${\bf E} = - \nabla \varphi$ in this case and the Lagrange multiplier $\varphi$ is the usual electric potential. 

We are aware that replacing the quasi-equilibrium with a typical monotonic approach to equilibrium will add a dipsersion to the impedance response. It is likely that multiexponential approaches to equilibrium, or overshoots, will produce the range of dispersions found in the impedance \cite{barsoukov2005impedance, banwell1972fundamentals,  gudarzi2021self, kremer2002broadband, rao2012molecular, raicu2015dielectric, sindhu2006fundamentals, steinfeld2012molecules, stuart2021infrared}.

For the mechanical part, a standard variational procedure (see Appendix A) leads to
\begin{equation}
  \begin{aligned}
\frac{k_B T}{D_i} c_i \uvec_i  & =  c_i \nabla \mu_i \\
& = k_B T \nabla c_i -  c_i \nabla \left(  {\textstyle \frac{1}{2}}  \tfrac{\pp  \varepsilon }{\pp c_i} |{\bf E}|^2  \right)  +  z_i q c_i \nabla \varphi\\
  \end{aligned}
\end{equation}
The final PNP equation with a concentration-dependent dielectric coefficient is given by
\begin{equation}
\begin{aligned}
  & \pp_t c_i = \nabla \cdot D_i \left( \nabla c_i + \tfrac{1}{k_B T} c_i \nabla \left( z_i q \varphi -  {\textstyle \frac{1}{2}}  \tfrac{\pp  \varepsilon }{\pp c_i} |\nabla \varphi|^2   \right)   \right) \\
  &  - \nabla \cdot \left( \varepsilon ({\bm c})\nabla \varphi ) \right) =  \sum_{i=1}^n q z_i c_i + \rho_0 (\x), \\
\end{aligned}
\end{equation}
The final equation is the same as (Eq. 40) of \cite{liu2018analysis} .

\subsection{PNP equation with an energy relaxation}


As mentioned earlier, $\rho_f(\x)$ may not be exactly the same as the mechanical charge $\hat{\rho}_f({\bm c})$. The difference may arise from charge displacement or a coarse graining procedure used in the analysis, e.g., if a stochastic term is involved, or atomic scale structures are present, coarse graining is hard to avoid as one reaches to mesoscopic and macroscopic scales of biological and technological systems.
Instead of using a Lagrange multiplier, which forces $\rho_f(\x)$ to be the same to $\hat{\rho}_f({\bm c})$,  we can use the following form of the free energy
\begin{equation}
  \begin{aligned}
    & \mathcal{F}({\bm c}, {\bm E})  = \int_{\Omega}  \sum_{i=1}^n c_i (\ln c_i - 1) + W_{\rm elec}({\bf E})  \\
    & \quad + \frac{M}{2}  \left( \nabla \cdot \left( \frac{\pp W_{\rm elec}}{\pp  {\bf E}} \right) -  \sum_{i=1}^n z_i c_i q - \rho(\x) \right)^2 \dd \x, \\
  \end{aligned}
\end{equation}
where the last term is the energy cost for the difference between mechanical charge density $\hat{\rho}_f({\bm c})$ and the electric charge density  $\rho_f(\x)$.


Again, we take $W_{\rm elec}({\bf E})$ as $W_{\rm elec}({\bf E}) = \frac{\varepsilon}{2} |{\bf E}|^2$ to illustrate the idea.
Similar to the previous calculations, in the case when the electrical part can immediately go to equilibrium, we have
\begin{equation}
\begin{aligned}
& \frac{\delta \mathcal{F}({\bm c}, {\bm E}) }{\delta {\bf E}} \\
& = \varepsilon {\bf E} - \varepsilon \nabla \left( M  \left( \nabla \cdot \left(  \varepsilon {\bf E} \right) -  \sum_{i=1}^n z_i c_i q - \rho_0(\x) \right) \right) = 0, \\
\end{aligned}
\end{equation}
which leads to
\begin{equation}\label{Def_E_M}
   {\bf E} = \nabla \left( M  \left( \nabla \cdot \left(  \varepsilon {\bf E} \right) -  \sum_{i=1}^n z_i c_i q - \rho(\x) \right) \right).
\end{equation}
Once again, we note that monotonic delays in reaching equilibrium are likely to produce the additional dispersions found throughout the literature of dielectric, impedance, and molecular spectroscopy \cite{barsoukov2005impedance, banwell1972fundamentals,  gudarzi2021self, kremer2002broadband, rao2012molecular, raicu2015dielectric, sindhu2006fundamentals, steinfeld2012molecules, stuart2021infrared}. Overshoot will lead to even more intriguing frequency responses and dispersions. 

Studying the dispersions found experimentally will include determining the energy functions, etc., needed to fit observed data. Note this is an inverse problem and must be approached as such because of the inherent ill-posed nature of inverse problems, whether presented as reverse engineering, parameter estimation, curve fitting, or inverse problem theory itself.

According to (\ref{Def_D_from_E}), we take 
\begin{equation}\label{eq_phi}
\varphi = - \left( M  \left( \nabla \cdot \left(  \varepsilon {\bf E} \right) -  \sum_{i=1}^n z_i c_i q - \rho(\x) \right) \right),
\end{equation}
which corresponds to usual definition of electrostatic potential formulations. From (\ref{eq_phi}), we can then obtain a equation for $\varphi$, given by
\begin{equation}
       - \nabla \cdot ({\varepsilon \nabla \varphi})  -  \sum_{i=1}^n z_i c_i q - \rho(\x) = \frac{- \varphi - C}{M}
\end{equation}
for any constant $C$ that can be taken as $0$.  Formally, we can recover the Poisson equation 
\begin{equation}
- \nabla \cdot (\varepsilon \nabla \varphi) =  \sum_{i=1}^n z_i c_i q + \rho(\x) 
\end{equation}
for the limit $M \rightarrow \infty$.

For the mechanical part, the chemical potential for the  $i$-th species can be computed as
\begin{equation}
  \begin{aligned}
\mu_i  & = \ln c_i - M  \left( \nabla \cdot (\varepsilon {\bf E}) - \sum_{i=1}^n c_i z_i q - \rho_0 (\x) \right) z_i q, \\
& = \ln c_i + \varphi z_i q, \\
  \end{aligned}
\end{equation}
which is exactly the same as the classical PNP case, although $\varphi$ no longer satisfies the Poisson equation for given $M$ with the free energy (\ref{Diss_PNP}). The final PNP system is given by 
\begin{equation}
  \begin{aligned}
    & \pp_t c_i = \nabla \cdot \left( D_i (\nabla c_i + \frac{q z_i}{k_B T} c_i \nabla \varphi ) \right) \\
    &  - \nabla \cdot ({\varepsilon \nabla \varphi})  =  \sum_{i=1}^n z_i c_i q + \rho(\x) + \frac{- \varphi}{M}. \\
  \end{aligned}
  \end{equation}
For more complicated the systems, other forms of energy relaxation can be used.

\section{Numerics: Current-voltage relation}
\begin{figure}[!h]
\centering
\includegraphics[width = 0.95 \linewidth]{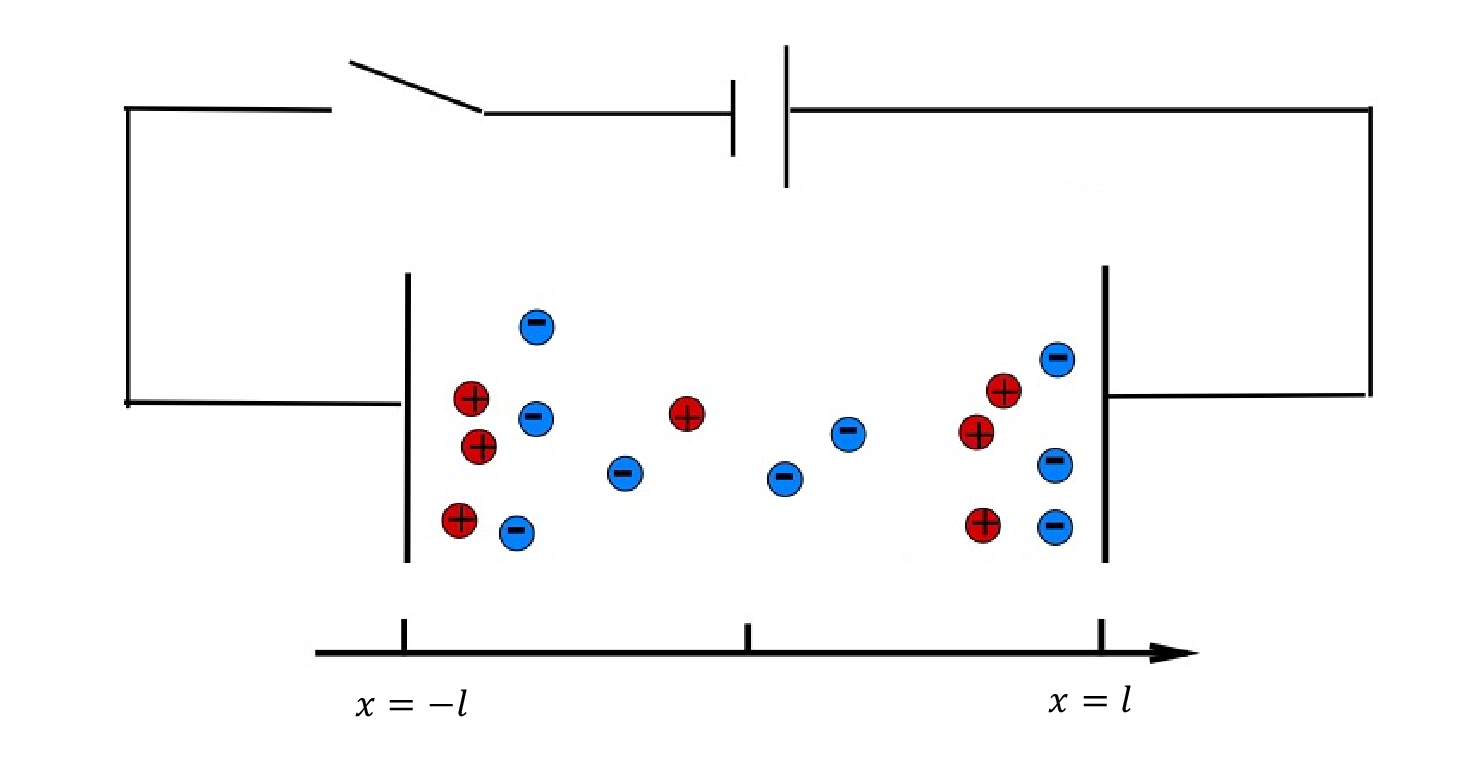}
\caption{A schematic illustration of a model system.}\label{Schem}
\end{figure}

In this section, we consider a model system, shown in Fig. \ref{Schem}. We study the current-voltage relation for the model system by applying 
a sinusoidal external potential in the tradition reaching back to the 1800's and the invention of the Wheatstone bridge.   

\begin{figure*}[!ht]
  \includegraphics[width = 0.95 \linewidth]{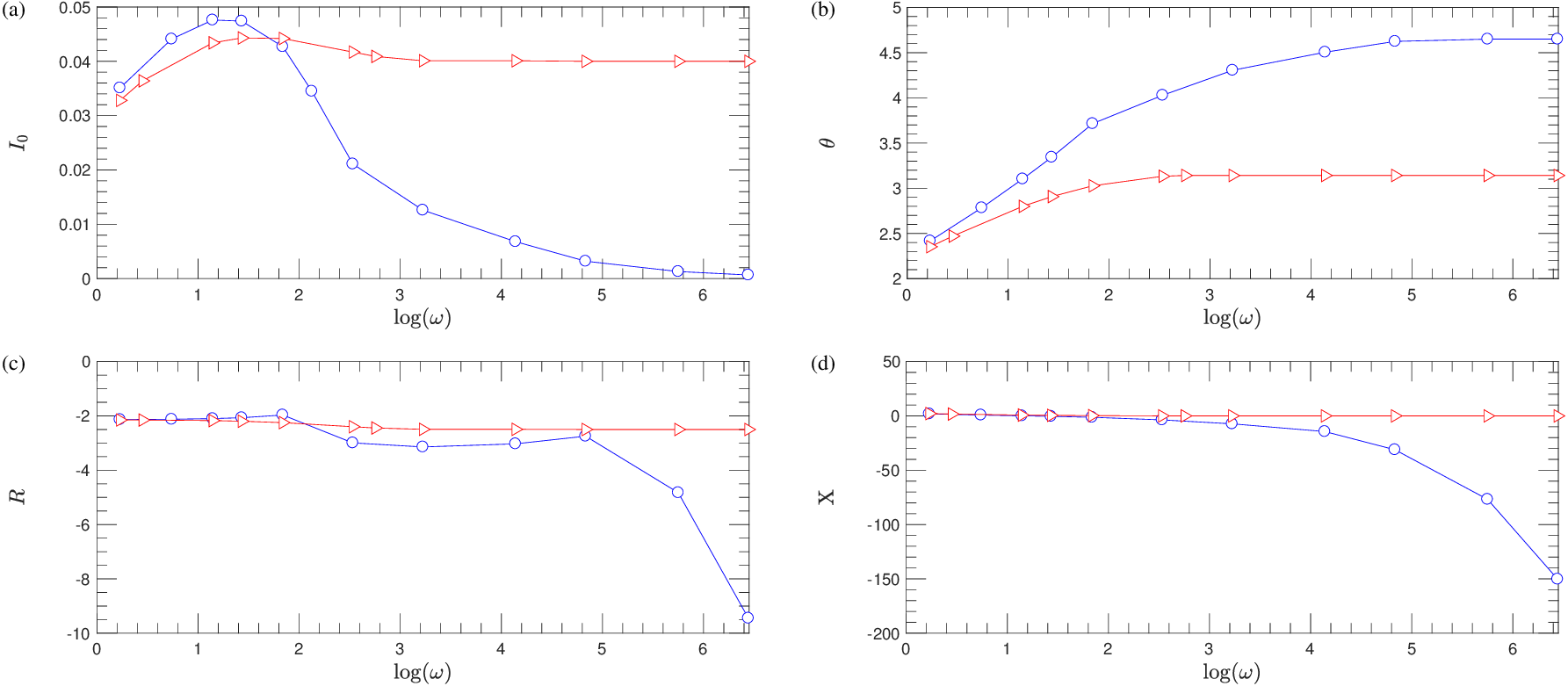}
\caption{Impedance plots for $\eta = 1$ [circle] and $\eta = 0$ [triangle] with $\varepsilon = 0.1$ with different $\omega$: (a) $\log(\omega)$ v.s. $I_0$, (b) $\log(\omega)$ v.s. $\theta$ , (c) $\log(\omega)$ v.s. $R = \frac{V_0}{I_0} \cos (\theta)$ , (d) $\log(\omega)$ v.s. $X = \frac{V_0}{I_0} \sin (\theta)$ }\label{Imp_plot}
\end{figure*}

The purpose of this section is to show that different energy functionals will lead to different impedance responses. The extensive literature on dielectric and impedance and molecular spectroscopy \cite{barsoukov2005impedance, banwell1972fundamentals,  gudarzi2021self, kremer2002broadband, rao2012molecular, raicu2015dielectric, sindhu2006fundamentals, steinfeld2012molecules, stuart2021infrared} thus can be used to determine the energy functions, and other parameters of our theory. 

The advantages of using a single energy based representation of these complex phenomena seem clear to us. The enormously valuable experimental literature can then be viewed in a single modern representation. Without this representation, it is possible that the experimental literature will fall away, out of sight. The understandably $ad$ $ hoc$ models of the classical literature are unfamiliar to our contemporaries---and their students and successors no doubt--and so a modern representation is needed to include classical experiments in future thinking, in our opinion.

It is important to reach to the older literature because it used sinusoidal analysis to provide detailed information about literally thousands of systems that remain of interest today. The advantages of the classical impedance spectra approach using a sinusoidal analysis for small signals is clear. It provides data that is most useful in determining the linear constitutive relations needed before more complex nonlinear properties are studied. Without the linear constitutive relations, it would be difficult, if not impossible to formulate the nonlinear relations in a well posed reasonably unique way. 

Frequency domain measurements play a special role in determining linear consititutive laws. The identification of underlying mechanisms is a great deal easier when the data is in the frequency domain.

Time domain measurements are not as helpful because of their huge dynamic range (because the underlying functions are exponentials that cannot be captured by ordinary electronic instrumentation) and the strong correlations between different data points, that make inverse problems particularly difficult to solve. In frequency domain analysis, the perfect correlations of neighboring time domain points are replaced in the frequency domain by  uncorrelated (actually orthogonal)  neighboring points. If the frequency domain points are determined by stochastic methods (e.g. that use sums of sinusoids with random phase as inputs), the points remain orthogonal and uncorrelated as discussed in textbooks of (digital and discrete) signal processing. 

As a generalization of the classical PNP systems, we consider a PNP system with inertial terms that introduce a delay in the approach to equilibrium.
More precisely, we consider a one-dimensional PNP equation with inertial term
  \begin{equation}
    \begin{aligned}
      & \pp_t n = - \pp_x (n u_n) \\
      & \eta (\pp_t u_n + u_n (\pp_x u_n)) + \frac{1}{D_n} u_n = - \pp_x \mu_n \\
      & \pp_t p = - \pp_x (n u_p) \\
      & \eta (\pp_t u_p + u_p (\pp_x u_p)) + \frac{1}{D_p} u_p = - \pp_x \mu_n \\
    \end{aligned}
  \end{equation}
  where $n(x)$ and $p(x)$ are concentrations for negative and positive ions respectively. The chemical potential $\mu_n$ and $\mu_p$ are given by
  \begin{equation}
  \begin{aligned}
    & \mu_n = \ln n - \phi(x)  
    & \mu_p = \ln p + \phi(x) 
  \end{aligned}
\end{equation}
and $\varphi$ satisfies the Poisson equation
\begin{equation}
  - \pp_x (\varepsilon(n, p) \pp_x \varphi(x)) = p - n,  
\end{equation}
The system can be reduced to a standard PNP equation when $\eta =  0$.
One can derive this system from the energy-dissipation law
\begin{equation}
\begin{aligned}
  & \frac{\dd}{\dd t} \int  \sum_{i=1}^2 \frac{\eta}{2} c_{i} |\uvec_{i}|^2 + \sum_{i=1}^2 K_B T c_i (\ln c_i - 1) + \frac{\varepsilon}{2} |{\bf E}|^2 \\
  & = - \int  \sum_{i=1}^2  \frac{1}{D_{i}} c_{i} |\uvec_{i}|^2 \dd \x,  \\
  \end{aligned}
  \end{equation}
  subject to the constraint
  \begin{equation}
     \nabla \cdot (\varepsilon {\bf E}) = c_2 - c_1,
  \end{equation}
  where $c_1 = n$ and $c_2 = p$.


To perform sinusoidal analysis, we impose a Dirichlet boundary condition for $\varphi$
\begin{equation}
  \begin{aligned}
  & \phi(\pm l, t) = \pm V_0  \cos (\omega t), \\   
  \end{aligned}
\end{equation}
where $\phi_0$ is the amplitude of the sinusoidal external potential, $f = \omega / 2 \pi$ is the frequency. Moreover, for the velocity field, we impose the boundary condition $u_n  = u_p = 0$. The initial condition is taken as
\begin{equation}
n(x, t) = 1, \quad p(x, t) = 1.
\end{equation}
The diffusion coefficients are taken as $D_n = D_p = 0.1$.
We look at current at $x = 0$. 
The current at a steady-state can be written as
\begin{equation}
 I(t) = I_0 \cos (\omega t - \theta(\omega)),
\end{equation}
where the phase angle $\theta$ depends on $\omega$. 

  In the complex-valued representation used widely in the classical literature, and throughout 
  electrical and electronic engineering,  
  \begin{equation}
    V = V_0 \exp(i (\omega t)), \quad I = I_0 \exp (i  (\omega t - \theta ))
  \end{equation}
  and the impedance is defined by
  \begin{equation}
Z = \frac{V}{I} = \frac{V_0}{I_0} \exp(i \theta) = R + i X,
  \end{equation}
where $R = \frac{V_0}{I_0} \cos \theta$ is the resistance, and  $X = \frac{V_0}{I_0} \sin \theta$ is the reactance. We plot $I_0$, $\theta$, $R$ and $X$ with respect $\log (\omega)$ in impedance plots. 
The impedance plots for $\eta = 1$ and $\eta = 0.1$ with $\varepsilon = 0.1$ and different $\omega$ are show in Fig. (\ref{Imp_plot}). 
One can build an analogy between a PNP system and a classical circuit or network \cite{kilic2007steric}.
In an LRC-series electric circuit, the current $i = \frac{\dd q}{\dd t}$ can be computed by solving the following ODE
\begin{equation}
  L \frac{\dd^2 q}{\dd t^2} + R \frac{\dd q}{\dd t} + \frac{1}{C} q = V_0 \cos \omega t,
\end{equation}
where $L$ is the inductance, $R$ is the resistance, $C$ is capacity, and $V_0 \cos \omega t$ is the applied voltage. The steady-state current can be computed as
steady-state is given by
\begin{equation}
  \begin{aligned}
i_p(t) 
                 & = \frac{V_0}{Z} \cos (\omega t - \alpha), \\
  \end{aligned}
\end{equation}
where $X = L \omega - \frac{1}{C \omega}$, $Z = \sqrt{X^2 + R^2}$, $\cos \alpha = \frac{R}{Z}$, and $\sin \alpha = \frac{X}{Z}$.
The impedance plots (\ref{Imp_plot}) suggest that the classical PNP system, without inertial terms, can be described by circuits without inductors, in which RC elements account for the time delays and dispersions. Inertial terms appear as inductors.

\section{Conclusion}
In this paper, we develop a new electric-field based variational formulation to model electromechanical systems. This framework is motivated and developed using the classical energetic variational approach (EnVarA). 
The dynamics and fluctuation are imposed in the mechanical part only. Imposing dynamics on the electrical part, or on both the electrical and mechanical parts of the system appears possible, but leads to complexities beyond the scope of this paper.

The coupling between the electric part and the chemo-mechanical part is described either by Lagrange multipliers or various energy relaxations. It is straightforward to extend the current formulation to non-isothermal cases and to systems that also involve chemical reactions as done by \cite{liu2020brinkman, wang2020field} in the EnVarA framework.
As an illustration, we re-derive the classical PNP system by both approaches and show the consistency of the current approach with the previous formulations. Numerical simulations show that different energy functionals (as estimated from experiments, for example) will lead to different impedance responses under a sinusoidal external potential. The form of the energy function can be sought then by solving an inverse problem, using additional structural information to reduce ill-posedness characteristic of inverse problems. The variational formulation can be applied more general electromechanical systems and opens a new door for developing structure-preserving numerical methods.

\section*{Acknowledge}
The authors acknowledge the partial support of NSF (Grant
No. DMS-1950868).

\appendix

\section{Derivation of (\ref{FB1})}
In this appendix, we give a detailed derivation of the force balance equation (\ref{FB1}) by the energetic variational approach. 

As mentioned earlier, a mechanical system is totally determined by the flow map $\x(\X, t)$ and the kinematics of the employed variables. Here $\X \in \Omega_0$ are Lagrangian coordinates and $\x \in \Omega$ are Eulerian coordinates. 

To apply the LAP, we need first reformulate the free energy in terms of the flow map $\x(\X, t)$. To this end, a Lagrangian description to the system is necessary. For a given flow $\x(\X, t)$, one can define the deformation tensor
\begin{equation}
{\sf F}(\X, t) = \nabla_{\X} \x(\X, t).
\end{equation}
Due to the conservation of mass, $c(\x, t)$ can be written as
\begin{equation}
c(\x(\X, t), t) = \frac{c_0(\X, t)}{\det {\sf F}}.
\end{equation}

As a consequence, the free energy can be reformulated as a functional of $\x(\X, t)$ in Lagrangian coordinates, i.e.,
\begin{equation}
\mathcal{F}[\x] = \int_{\Omega_0} \omega\left(\frac{c_0(\X)}{\det {\sf F}} \right) \det {\sf F} \, \dd \X.
\end{equation}
Then we can compute the variation of $\mathcal{A} = \int_{0}^{T} \int \omega \dd \x $ with respect to $\x(\X, t)$. 

Indeed, a direct computation leads to
\begin{equation*}
  \begin{aligned}
    & \delta \mathcal{A} = - \delta \int_{0}^T  \int_{\Omega_0} \omega(c_0(X)/ \det F) \det F \, \dd \X \\
    & = - \int_{0}^T \int_{\Omega_0} \left( - \frac{\pp \omega}{\pp c}  \left( \frac{c_0(X)}{\det F} \right)  \cdot \frac{c_0(X)}{\det F} + \omega\left(\frac{c_0(X)}{\det F}\right) \right) \\
    & \quad \quad \quad \quad  \quad \quad  \times  (F^{-\rm{T}} : \nabla_{\X} \delta \x)\det F \, \, \dd \X, \\
      \end{aligned}
\end{equation*}
where $\delta \x$ is the test function satisfying $\tilde{\delta \x} \cdot \n = 0$ and $\n$ is the outer normal of $\Omega$. When we pull back to Eulerian coordinate, we have
\begin{equation}
  \begin{aligned}
\delta \mathcal{A} & = - \int_{0}^T \int_{\Omega} ( -  \frac{\pp \omega}{\pp c} c + \omega) \nabla \cdot (\delta \x) \dd \x \\
& = \int_{0}^T \int_{\Omega} - \nabla  (\frac{\pp \omega}{\pp c} c - \omega) \cdot \delta \x  \dd \x \\
  \end{aligned} 
\end{equation}
Hence, $$\frac{\delta \mathcal{A}}{\delta \x} = - \nabla (\frac{\pp \omega}{\pp c} c - \omega) = - c \nabla \mu,$$ where $\mu = \frac{\pp \omega}{\pp c}$ is the chemical potential. For the dissipation part, since $\mathcal{D} = \frac{1}{2} \int  \eta(c) |\x_t|^2  \dd \x$ it is easy to compute that $\frac{\delta \mathcal{D}}{\delta \x_t} = \eta(c) \x_t$.



\bibliography{PNP}

\end{document}